\documentclass[pre,amsmath,showpacs]{revtex4}
\usepackage{graphicx}
\begin{document}
\title{Wetting under non-equilibrium conditions}
\author{H. Hinrichsen$^{1,2}$, R. Livi$^{3}$,
    D. Mukamel$^{1}$, and A. Politi$^{4}$}
\affiliation{$^{1}$ Department of Physics of Complex Systems,
             Weizmann Institute of Science, Rehovot 76100, Israel}
\affiliation{$^{2}$ Theoretische Physik, Fachbereich 8,
             Bergische Universit{\"a}t Wuppertal, 42097 Wuppertal, Germany}
\affiliation{$^{3}$ Dipartimento di Fisica, Universit\'a, INFM and INFN,
             50125 Firenze, Italy}
\affiliation{$^{4}$ Istituto Nazionale di Ottica Applicata and INFM-Firenze,
             50125 Firenze, Italy}

\begin{abstract}
We report a detailed account of the phase diagram of a recently
introduced model for non-equilibrium wetting in $1+1$ dimensions
[Phys. Rev. Lett. {\bf 79}, 2710 (1997)]. A mean field
approximation is shown to reproduce the main features of the phase
diagram, while providing indications for the behaviour of the
wetting transition in higher dimensions. The mean field phase
diagram is found to exhibit an extra transition line which does
not exist in $1+1$ dimensions. The line separates a phase in which
the interface height distribution decays exponentially at large
heights, from a superexponentially decaying phase. Implications to
wetting in dimensions higher than $1+1$ are discussed.

\end{abstract}


\maketitle
\def\makevector#1{\text{\bf#1}}
\def\xvec{\makevector{x}}
\parskip 2mm

\section{Introduction}

Wetting phenomena occur in a large variety of experiments, where a
planar substrate is exposed to a gas phase under thermal
equilibrium conditions. Generally, `wetting' refers to a situation
where a bulk phase $(\alpha)$ in contact with a substrate coexists
with a layer of a different phase ($\beta$) which is
preferentially attracted to the surface of the substrate. By
changing physical parameters such as temperature and chemical
potential, the system may undergo a wetting transition from a
non-wet phase, where the thickness of the layer stays finite, to a
wet phase, where the layer becomes macroscopic.

The phase diagram associated with the surface layer could be
rather complex exhibiting a variety of surface phase transitions,
prewetting phenomena and multicritical
behavior~\cite{WettingReview,NakanishiFisher}. For example, by
increasing the temperature $T$ while moving along the
$\alpha$-$\beta$ coexistence curve, a wetting transition may take
place at a temperature $T_W$, beyond which the thickness of the
layer becomes infinite. Usually this transition is first order,
although in certain models the transition is continuous, and is
then referred to as continuous wetting. On the other hand, when
the chemical potential difference between the two phases is
varied, moving towards the coexistence curve at $T > T_W$, a
different type of transition takes place in which the thickness of
the layer diverges. This phenomenon is known as complete wetting.

In many experimental situations it is reasonable to assume that
a wetting stationary layer is in thermal equilibrium. In fact,
methods of equilibrium statistical mechanics turned out to be very
successful in a large variety of theoretical and experimental
studies (for a review, see Ref.~\cite{WettingReview}). Within this
approach, a wetting transition is usually considered as the
unbinding of an interface from a wall. The interface configuration
is described by a function $h(\xvec)$ which gives the height of
the interface at point $\xvec$ on the substrate.
One then introduces an effective Hamiltonian of the
form~\cite{EquilibriumFieldTheory}
\begin{equation}
\label{Hamiltonian}
{\cal H} = \int d^{d-1}x \biggl[ \frac{\sigma}{2}
(\nabla h)^2 + V\bigl(h(\xvec)\bigr) \biggr] \,,
\end{equation}
where $\sigma$ is the effective surface tension of the
$\alpha$-$\beta$ interface, $V(h)$ is a potential
accounting for the interaction between the wall and the interface,
and $d-1$ is the interface dimension. In the non-wet phase the
potential $V$ contains an attractive component which binds the
interface to the wall. Assuming thermal equilibrium, the
probability of finding the interface in a certain configuration is
then given by the canonical distribution
\begin{equation}
\label{EquilibriumEnsemble}
P[h] \sim \exp\bigl(-\beta {\cal H}[h]\bigr).
\end{equation}
As the parameters describing the system are varied, the attractive
component of the potential may become weaker so that it is no
longer able to bind the interface, leading to a wetting
transition.

In order to study wetting phenomena under thermal equilibrium
conditions one usually introduces a stochastic Langevin equation
corresponding to the effective Hamiltonian~(\ref{Hamiltonian}).
This Langevin dynamics should reproduce the equilibrium
distribution~(\ref{EquilibriumEnsemble}) in the limit $t \to
\infty$. Since many different dynamical rules may approach the
same stationary state, this condition does not fully determine the
form of the Langevin equation. However, assuming short-range
interactions and keeping only the most relevant terms in the
renormalization group sense, one is led to the Edwards-Wilkinson
equation with a potential~\cite{BarabasiStanley}
\begin{equation}
\label{EW}
\frac{\partial h(\xvec,t)}{\partial t} = \sigma\nabla^2 h(\xvec,t) -
\frac{\partial V\bigl(h(\xvec,t)\bigr)}{\partial h(\xvec,t)}
 \noindent + \zeta(\xvec,t)\,,
\end{equation}
where $\zeta(\xvec,t)$ is a zero-average Gaussian noise field with
a variance
\begin{equation}
\label{Noise} \langle\zeta(\xvec,t)\zeta(\xvec',t')\rangle=
2\Gamma\delta^{d-1}(\xvec-\xvec')\delta(t-t')\,,
\end{equation}
and a noise amplitude $\Gamma=k_BT$. This Langevin
equation has the same symmetry properties as the
Hamiltonian~(\ref{Hamiltonian}), namely, it is invariant under
translations, rotations, and reflections in space. Apart from the
potential term, the equation is also invariant under shifts $h\to
h+a$ and reflections $h \to -h$. Moreover, it can be shown that
this type of Langevin dynamics obeys {\em detailed balance} and
relaxes towards the equilibrium
distribution~(\ref{EquilibriumEnsemble}) in the bound phase.

Wetting phenomena may also take place in many systems under
non-equilibrium conditions. For example, in growth processes such
as molecular beam epitaxy or others, a layer is grown on a
substrate, whose properties depend on the growth conditions. By
varying these conditions one expects wetting phenomena to take
place. Here, unlike the equilibrium case, the dynamics does not
obey detailed balance, leading to a rather different class of
wetting phenomena.

The simplest way to study nonequilibrium wetting on the level of
the Langevin equation is to introduce a nonlinear term in
Eq.~(\ref{EW}), leading to a Kardar-Parisi-Zhang (KPZ) equation
with a potential~\cite{KPZWall}
\begin{equation}
\label{LangevinEquation}
\frac{\partial h(\xvec,t)}{\partial t} = \sigma\nabla^2 h(\xvec,t) -
\frac{\partial V\bigl(h(\xvec,t)\bigr)}{\partial h(\xvec,t)}
 \noindent + \lambda \bigl(\nabla h(\xvec,t)\bigr)^2 + \zeta(\xvec,t)\,.
\end{equation}
It is important to note that this nonlinear term is a relevant perturbation of
the underlying field theory, i.e., even if $\lambda$ is very small,
it will be amplified under renormalization group transformations, driving the system
away from thermal equilibrium.

Recently, a simple solid-on-solid (SOS) model for non-equilibrium
wetting in $1+1$ dimensions was introduced~\cite{Alon,Wetting1}.
The model is controlled by an adsorption rate $q$ and a desorption
rate $p$ and exhibits a continuous wetting transition at a
critical growth rate $q_c(p)$. The wetting transition is related
to the unpinning process of an interface from a substrate and may
be described by the KPZ equation~(\ref{LangevinEquation}). The
model has then been generalized to include a short-range
interaction between the interface and the
substrate~\cite{Wetting2}. This was done by introducing a modified
growth rate $q_0$ at the substrate level. This results in a
contact interaction between the interface and the substrate, which
is attractive for $q_0 < q$ and repulsive for $q_0 > q$. It was
found that sufficiently strong attractive interaction modifies the
nature of the wetting transition, making it first order. In
addition, it has been demonstrated that there exists an extended
region in the phase diagram, where the pinned and the moving
phases {\em coexist} in the sense that the transition time from
the pinned to the moving phase grows exponentially with the
system size so that the two phases become stable in the
thermodynamic limit. It should be emphasized that this kind of
phase coexistence, which has also been observed in the past in
other models~\cite{Toom80,BennettGrinstein85,Bridge95,Mueller97},
can only occur in nonequilibrium systems.

Some of these results have been confirmed by numerical and mean
field studies of KPZ type models~\cite{GiadaMarsili,Santos}. In
particular, by adding a repulsive interaction between the
interface and the substrate, a crossover from a continuous to a
first order wetting transition was found. Non-equilibrium wetting
phenomena have also been studied recently in models of growing
magnetic domains~\cite{CandiaAlbano}.

In this paper we present a detailed account of the phase diagram
of the SOS model in 1+1 dimensions. In order to get some
indication on the behavior of the model in higher dimensions we
introduce a mean field approximation for the model, and study the
resulting phase diagram and the dynamical behavior of the
interface. It is found that the mean field approximation
reproduces the main qualitative features of the phase diagram
obtained in 1+1 dimensions. In addition, a new feature is found in
the pinned phase. By studying the height distribution of the
pinned interface, two distinct types of behavior are discovered
upon varying the dynamical parameters of the model: in one regime
the distribution decays superexponentially with the height, while
in the other the decay is exponential. The two regimes are
separated by a prewetting transition line. Relevance of this
phase diagram to wetting phenomena in dimensions higher than $1+1$
is considered.

The paper is organized as follows. In Sec.~\ref{DefinitionSection}
we recall the definition of the SOS model and summarize its
properties. The special case $p=1$, where the model is exactly
soluble, is discussed in detail in Sec.~\ref{SecDetailedBalance}.
The mean field approximation, which is the main focus of
the present work, is presented and analyzed in
Sec.~\ref{MFSection}. The paper ends with concluding remarks in
Sec.~\ref{SecConcl}.

\section{Definition and properties of the model}
\label{DefinitionSection}
%
%
\begin{figure}
\includegraphics[width=90mm]{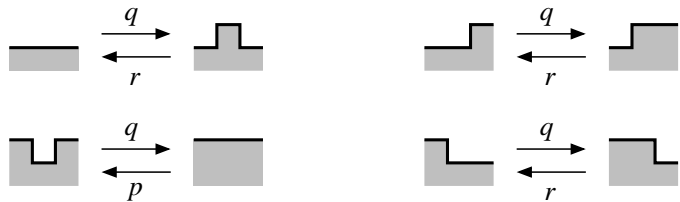}
\caption{\footnotesize
\label{FIGRULES}
Deposition and evaporation processes in the bulk. At the
hard-core wall at zero height (not shown here) evaporation is forbidden and
the deposition rate $q$ is replaced by a modified deposition
rate $q_0$ in order to take the interaction between substrate
and surface layer into account.
}
\end{figure}
The SOS model defined in~\cite{Wetting1} is probably the simplest
model which follows the spirit of Eq.~(\ref{LangevinEquation}). It
is defined on a one-dimensional lattice with $N$ sites and
periodic boundary conditions, where each site~$i$ is associated
with an integer variable $h_i$ describing the local height of the
interface. The repulsive part of the potential $V(h)$ is
implemented as a hard-core wall at zero height, restricting the
heights $h_i$ to be non-negative. Thus $h_i=0,1,2,\dots,$ .
Moreover, an effective surface tension is introduced by imposing
the restricted solid-on-solid (RSOS) condition
\begin{equation}
\label{RSOS}
|h_i-h_{i\pm 1}| \leq 1\,.
\end{equation}
The model evolves random-sequentially by choosing a random
site and carrying out one of the following processes
(see Fig.~\ref{FIGRULES}):
\begin{itemize}

\item[-] Deposition of an atom on the substrate with rate $q_0$:
        \begin{equation} \label{Process0}
        h_i=0 \rightarrow h_i=1 \end{equation}

\item[-] Deposition of an atom on top of already deposited islands with rate $q$:
        \begin{equation} \label{Process1}
        h_i \rightarrow h_i+1 \qquad \text{if }h_i \geq 1\end{equation}

\item[-] Evaporation of an atom at the edge of a terrace with rate $r$:
        \begin{equation} \label{Process2}
        h_i \rightarrow \min(h_{i-1},h_{i},h_{i+1})
        \end{equation}

\item[-] Evaporation of an atom in the middle of a plateau
        with rate $p$:
        \begin{equation}  \label{Process3}
        h_i \rightarrow h_i-1 \quad \mbox{if} \quad
        h_{i-1}=h_{i}=h_{i+1}>0 \end{equation}

\end{itemize}
If the selected process would violate the restrictions $h_i\geq 0$
or $|h_i-h_{i\pm 1}| \leq 1$, the attempted move is abandoned and
a new site $i$ is selected. Each attempted update corresponds to a
time increment $\Delta t=1/N$. Since one of the four rates can be
chosen freely by rescaling time, we set $r=1$. Thus the model is
controlled by three parameters, namely, a growth rate $q$, a
desorption rate $p$, and a special growth rate $q_0$ at the
substrate which accounts for an additional short-range interaction
between the substrate and the wetting layer.

The model can be easily generalized to higher dimensions. Note
that in this case it is possible to introduce different
evaporation rates for various types of edge sites, e.g. linear
edges and corner sites. For simplicity we will assume that all
rates for evaporation at edges are equal to $1$. In this case the
model can simply be generalized to higher dimensions by including
all nearest neighbors in Eqs.~(\ref{Process2})
and~(\ref{Process3}).

\subsection{Properties for $q_0=q$}

\begin{figure}
\includegraphics[width=80mm]{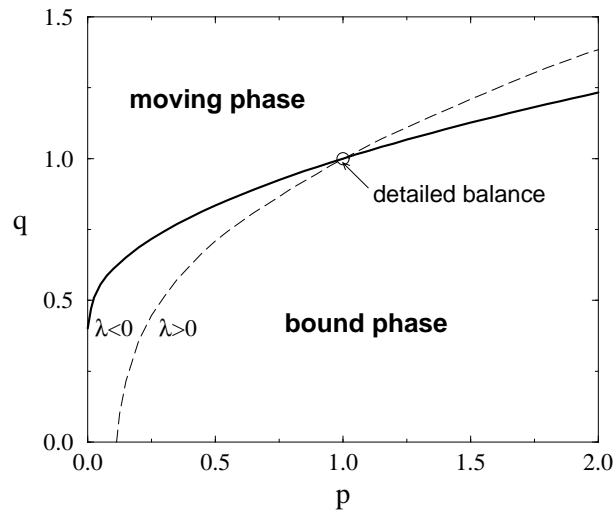}
\caption{\footnotesize \label{FIGPHASETRANS} Phase diagram of the
wetting model for $q_0=q$. The second-order wetting transition is
represented as a solid line. The dashed line indicates where the
coefficient $\lambda$ of the nonlinear term in the KPZ equation
effectively vanishes. For $p=1$ the dynamical rules obey detailed
balance (see text). }
\end{figure}

Let us first consider the case without interactions between
substrate and wetting layer, i.e., $q_0=q$. In this case the presence
of a hard-core wall at zero height leads to a {\em continuous} phase
transition between a bound and a moving phase (see Fig.~\ref{FIGPHASETRANS}).
The phase transition line $q_c(p)$ is determined by a vanishing propagation
velocity of a freely evolving interface. For $q<q_c$ the interface moves
downward until it fluctuates close to the wall while for $q>q_c$
the propagation velocity is positive and the interface detaches
from the wall. Obviously this transition takes place even
in finite systems with a critical threshold depending on $N$. 
We note that for $p=0$, where evaporation
from the middle of plateaus is forbidden, the interface velocity
in the bulk cannot be negative so that in this case
the transition relies on a different mechanism~\cite{Alon}.

\begin{figure}
\includegraphics[width=90mm]{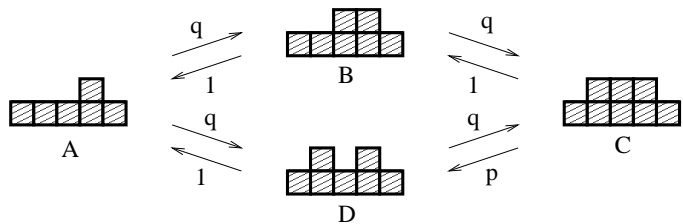}
\caption{\footnotesize
\label{FIGDB}
Example of a closed cycle of deposition and evaporation processes,
showing that detailed balance can only be established if $p=1$.
}
\end{figure}

One can easily verify that the dynamical rules
(\ref{Process0})-(\ref{Process3}) do not generally satisfy
detailed balance. To this end we consider the closed cycle of
deposition and evaporation processes, as shown in
Fig.~\ref{FIGDB}. Let the statistical weights of these
configurations in the steady state be $P_A,P_B,P_C$, and $P_D$.
Obviously detailed balance implies
\begin{equation}
q^2 P_A = q P_B = P_C =\frac{q}{p}P_D = \frac{q^2}{p}P_A \,.
\end{equation}
These equations can only be satisfied if $p=1$ (as already
anticipated, $r$ is assumed to be equal to 1 without any loss of
generality). This special case will be discussed in more detail in
Sec.~\ref{SecDetailedBalance}.

The two parameters $p$ and $q$ can be used to control $\partial V/\partial h$
and $\lambda$ in the KPZ equation~(\ref{LangevinEquation}).
In the case of detailed balance, where a bound interface is
thermally equilibrated, the coefficient $\lambda$ is expected
to vanish. This can be verified by comparing the propagation velocities
of a horizontal and an artificially tilted interface far away from the wall.
The line, where both velocities coincide, is shown as a dashed line in 
Fig.~\ref{FIGPHASETRANS}. As expected, it crosses the phase transition
line exactly at the point $p=q=1$, where detailed balance is satisfied.

Moving away from this point, we can therefore study the crossover from
equilibrium to non-equilibrium wetting. Numerical simulations
suggested that the transitions for $p<1$ and $p>1$ are associated
with different sets of critical exponents~\cite{Wetting1}.
These findings are in accordance with results obtained by Mun\~oz and
Hwa~\cite{KPZWall}, who showed that the scaling properties of a KPZ
interface interacting with a wall depend on the sign of $\lambda$.

\subsection{Properties for $q_0<q$}

%
%
\begin{figure}
\includegraphics[width=80mm]{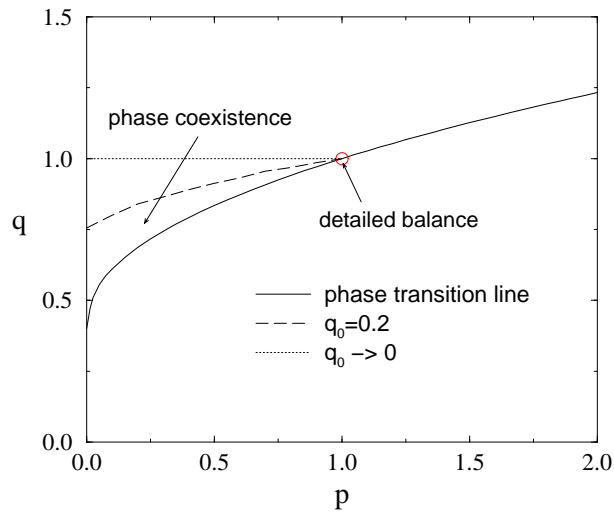}
\caption{\footnotesize
\label{FIGCOEX}
Phase coexistence region in the phase diagram for small values of $q_0$.
}
\end{figure}

In most experimental applications the wetting layer interacts with
the substrate, giving rise to an additional short-range force at
the bottom layer which may be either attractive or repulsive. In
the present model such a force can be taken into account by
introducing a different growth rate $q_0$ for deposition at zero
height. The influence of this parameter was studied in detail in
Ref.~\cite{Wetting2}. Since $q_0$ does not influence the
propagation velocity of the interface far away from the wall, the
location of the transition line, along which the moving phase is
stable, remains unchanged. However, if $q_0$ is smaller than a
certain threshold $q_0^*$, the attractive interaction is so strong
as to stabilize the bound phase even when the free interface would
grow. In other words, for very low values of $q_0$, there exists an
extended region in the phase diagram where the bound and the
moving phase {\em coexist} in the sense that the transition time
from the bound to the moving phase grows exponentially with the
systems size. The upper boundary of the coexistence region depends
on $q_0$, as shown in Fig.~\ref{FIGCOEX}.

A thermodynamically stable coexistence of the bound and the moving
phase requires a robust mechanism which eliminates large
protruding islands in the bound phase. In the present model this
mechanism works as follows. Once an island has been formed by
fluctuations, the detached part of the interface quickly grows
since $q>q_c$. In the phase coexistence region, where the
coefficient $\lambda$ in the KPZ equation is negative, the island
will grow until the slope at the edges exceeds a critical value,
where the growth is compensated by the nonlinear term. Afterwards,
the pyramidial island shrinks linearly with time until it is
eliminated. Therefore, phase coexistence can only occur under
non-equilibrium conditions in those regions of the phase diagram
where $\lambda$ is negative. Very recently, the tricritical point
and the critical behavior at the upper boundary of the phase 
coexistence region has been investigated in a discretized
KPZ equation with a potential~\cite{Ginelli}. 

\section{Exactly soluble case: $p=1$}
\label{SecDetailedBalance}

\subsection{Detailed balance and transfer matrix approach}

\begin{figure}
\includegraphics[width=70mm]{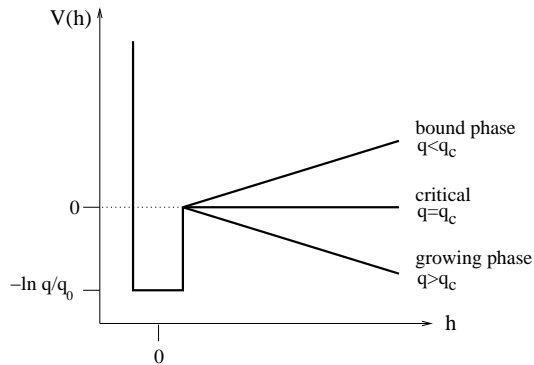}
\caption{\footnotesize
\label{FIGPOTENTIAL}
Schematic drawing of $V(h)$ with a potential well at zero height.
}
\end{figure}

We first consider the special case $p=1$, where detailed balance
is satisfied. In this case the stationary probability distribution
of the bound interface configurations $\{h_1,\dots,h_N\}$ is given
by the canonical ensemble expression corresponding to an energy
functional ${\cal H}$
\begin{equation}
\label{EquilibriumDistribution} P(h_1,\ldots,h_N) = \frac{1}{Z_N}
\exp\bigl[-{\cal H}(h_1,\ldots,h_N)\bigr] \ .
\end{equation}
The partition sum
\begin{equation}
Z_N=\sum_{h_1,\ldots,h_N} \exp\bigl[-{\cal H}(h_1,\ldots,h_N)\bigr]
\end{equation}
runs over all interface configurations
obeying the RSOS constraint~(\ref{RSOS}).
The energy ${\cal H}$ is given by
\begin{equation}
{\cal H}(h_1,\ldots,h_N)=\sum_{i=1}^N V(h_i) \,,
\end{equation}
where $V(h)$ is a potential of the form
\begin{equation}
\label{Potential}
V(h)=\left\{
\begin{array}{ll}
\infty      & \mbox{if \ } h<0  \\
-\ln(q/q_0)     & \mbox{if \ } h=0  \\
-h \; \ln(q)    & \mbox{if \ } h>0
\end{array}
\right.
\end{equation}
as sketched in Fig.~\ref{FIGPOTENTIAL}.
Therefore, Eq.~(\ref{EquilibriumDistribution}) may be rewritten as
\begin{equation}
\label{EquilibriumDistribution2}
P(h_1,\ldots,h_N)=Z_N^{-1} q^{(\sum_{i=1}^Nh_i)} \,
(q/q_0)^{(\sum_{i=1}^N\delta_{h_i,0})}\,.
\end{equation}
Note that in this expression
$\ln(q)$ plays the role of the chemical potential difference $\Delta \mu/k_BT$
between the wetting layer and the gas phase. Obviously the transition takes
place at $q_c=1$.

To verify the validity of this probability distribution, it is
sufficient to demonstrate that the dynamical rules obey detailed
balance with respect to it. In fact, the deposition
process~(\ref{Process1}) reduces the probability $P$ by a factor
of $q$ while the evaporation processes (\ref{Process2}) and
(\ref{Process3}), which both take place with the same rate,
increase $P$ by a factor of $1/q$. Consequently, the probability
currents between pairs of configurations compensate each other so
that detailed balance is satisfied, proving the validity of the
equilibrium ensemble~(\ref{EquilibriumDistribution2}). Similarly
one can show that detailed balance is also satisfied at the bottom
layer. We note that these considerations are only valid in the
bound phase $q<1$, where the probability distribution is
stationary. Once the system enters the moving phase, the process
is out of equilibrium.

The canonical ensemble can be used to compute the density profile
of a bound interface in the case of detailed balance. To this end
we use a transfer matrix formalism introduced
in~\cite{Hilhorst,Burkhardt}, writing the Boltzmann factor
$\exp(-{\cal H})$ in Eqs.~(\ref{EquilibriumDistribution})
and~(\ref{EquilibriumDistribution2}) as a product
\begin{equation}
\label{EquilibriumDistribution3}
P(h_1,\ldots,h_N)=\frac{1}{Z_N}
\prod_{i=1}^N T_{h_i,h_{i+1}}\,,
\end{equation}
where $Z_N = \text{Tr} ( T^N)$ and
\begin{equation}
T_{h,h'} =
\begin{cases}
q^{(h+h')/2} (q/q_0)^{(\delta_{h,0}+\delta_{h',0})/2} &
\text{ if } |h-h'| \leq 1 \\
0 &   \text{ otherwise. }
\end{cases}
\end{equation}
The transfer matrix $T$ is infinite-dimensional,
acts in spatial direction, and yields the
contribution to the Boltzmann factor between
adjacent sites with the heights $h$ and~$h'$.
Because of the RSOS condition~(\ref{RSOS}),
it has a tridiagonal structure and reads
\begin{equation}
\label{Tmatrix}
T=
\begin{pmatrix}
q/q_0 & q/q_0^{1/2} &&&&& \\
q/q_0^{1/2} & q & q^{3/2} &&&& \\
& q^{3/2} & q^2 & q^{5/2} &&& \\
&& q^{5/2} & q^3 & q^{7/2} && \\
&&& \ldots & \ldots & \ldots & \\
&&&& \ldots & \ldots & \ldots
\end{pmatrix} \,.
\end{equation}
Using the transfer matrix formalism,
the stationary density $\rho(h)$
of sites at height $h$ can be expressed as
\begin{equation}
\rho(h) = Z_N^{-1} \langle h |T^N | h \rangle \,,
\end{equation}
where $\{|h\rangle\}$, $\{\langle h |\}$
denote canonical basis vectors in height space.
For $N \to \infty$ this expression is governed by the
largest eigenvalue $\Lambda$ of the transfer matrix
\begin{equation}
\label{EigenvalueProblem}
T |\phi\rangle = \Lambda |\phi\rangle \,,
\end{equation}
where $|\phi\rangle$ denotes the corresponding eigenvector.
Thus, in an infinite system, the stationary densities may also be written as
\begin{equation}
\label{StationaryDens}
\rho(h) = \frac{|\langle h | \phi \rangle|^2}
{\langle \phi | \phi \rangle} \,.
\end{equation}
Note that the numerator in this expression is quadratic in
$|\phi\rangle$, just as in a quantum-mechanical problem.

\subsection{The case $q_0=q$}
\label{SecondOrderScaling}

In order to understand how the interface detaches from the wall,
it is useful to study the scaling behavior of the density of sites
at the bottom layer close to the transition point. Let us first
consider the case $q_0=q$, where interactions between substrate
and bottom layer are absent. For $h>0$ the eigenvalue problem reads
\begin{equation}
\label{EigevEquation}
q^{h-1/2}\phi_{h-1} + q^h\phi_{h} + q^{h+1/2}\phi_{h+1} = \Lambda \phi_{h} \,,
\end{equation}
where $\Lambda$ is the largest eigenvalue of $T$ and $\phi_h$ are
the components of the corresponding eigenvector $|\phi\rangle$
representing the stationary state. Close to criticality we can
carry out the continuum limit $\phi_h \rightarrow
\phi(\tilde{h})$, replacing discrete heights~$h$ by real-valued
heights~$\tilde{h}$. In this limit the above eigenvalue problem
turns into a differential equation which, to leading order in
$\epsilon=1-q >0$, is given by
\begin{equation}
\label{DifferentialEquation}
\Bigl( \frac{\partial^2}{\partial \tilde{h}^2} + (3-\Lambda) -
3 \epsilon \tilde{h} \Bigr)\, \phi(\tilde{h}) = 0\,.
\end{equation}
This differential equation is solved by an Airy function
\begin{equation}
\phi(\tilde{h}) =
\text{Ai} \biggl( \frac{3 \epsilon \tilde{h} + \Lambda - 3}
{(3\epsilon)^{2/3}} \biggr) \,.
\end{equation}
Since $\phi(\tilde{h})$ has to vanish for $\tilde{h}\leq 0$,
we obtain the eigenvalue $\Lambda=3$ so that $\phi(\tilde{h}) =
\text{Ai}([3\epsilon]^{1/3}\tilde{h})$. Therefore, the heights,
in particular the average height and the interface width,
scale as
\begin{equation}
\label{HScaling}
\langle \tilde{h} \rangle \sim w \sim
\epsilon^{-1/3}.
\end{equation}
Next, we determine the density of exposed sites at the
substrate $\rho(0)$. In the continuum limit of Eq.~(\ref{StationaryDens})
$\langle 0 | \phi \rangle$ is proportional to $\phi'(0)$, hence
\begin{equation}
\rho(0) = \frac{1}{\cal N} [\phi'(0)]^2\,,
\end{equation}
where ${\cal N} = \int_0^\infty d\tilde{h}\, \phi^2(\tilde{h})$ is a
normalization factor. Since $\phi'(0) \sim \epsilon^{1/3}$ and
${\cal N} \sim \epsilon^{-1/3}$ one obtains a linear scaling law
\begin{equation}
\label{LinearRho}
\rho(0) \sim \epsilon \,.
\end{equation}
Thus the density of exposed sites at the bottom layer scales
{\em linearly} with the distance from criticality, proving that
the wetting transition at $p=q_c=1$ is continuous.

\begin{figure}
\includegraphics[width=60mm]{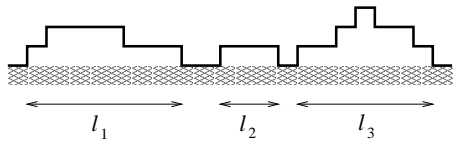}
\caption{\footnotesize
\label{FIGISLANDS}
1+1-dimensional interface with island sizes $\ell_1, \ell_2, \ell_3$.
}
\end{figure}

Imposing fixed boundary conditions $h_1=h_N=0$, it is also
possible to study finite-size scaling at the critical point
$q_0=q=1$. At criticality the transfer matrix has a simple structure
and can be thought of as generating a
simple random walk near a wall so that the mean height and
the bottom layer density scale as
\begin{equation}
\label{FSScaling}
\langle \tilde{h} \rangle \sim N^{1/2}\,, \qquad
\rho(0) \sim N^{-3/2}\,.
\end{equation}
The transfer matrix does not provide any information regarding
dynamical properties. Numerical simulations (whose details are not
shown here) suggest that an initially flat interface at zero
height roughens with time (for $t \ll N^2$) as
\begin{equation}
\label{TimeScaling}
\langle \tilde{h} \rangle \sim t^{1/4}\,,
\qquad \rho(0) \sim t^{-3/4}\,.
\end{equation}
Assuming standard power law scaling, we can combine Eqs.~(\ref{HScaling}),
(\ref{LinearRho}), (\ref{FSScaling}), and (\ref{TimeScaling})
in the scaling forms
\begin{equation}
\begin{split}
\langle \tilde{h}(\epsilon,N,t) \rangle &= N^{1/2} f\Bigl(t/N^2,\epsilon N^{3/2}\Bigr)\\
\rho_0(\epsilon,N,t)    &= N^{-3/2} g\Bigl(t/N^2,\epsilon N^{3/2}\Bigr)
\end{split}
\end{equation}
where $f$ and $g$ are scaling functions with an appropriate
asymptotic behavior. As expected these scaling forms are
consistent with the critical exponents $z=2, \alpha=1/2$ of the
Edwards-Wilkinson universality class~\cite{BarabasiStanley}.


Another interesting aspect is the stationary
distribution $P(\ell)$ of island sizes
$\ell$ in the bound phase (see Fig.~\ref{FIGISLANDS}).
In the case of detailed balance this distribution can be
computed exactly. To this end we introduce the
projection operator $P=1-|0\rangle\langle0|$ which
projects onto states with nonzero height. Moreover, let
$Q=PTP$ be a transfer matrix describing an interface that
does not touch the bottom layer. Obviously the distribution $P(\ell)$,
which may be interpreted as a first-return probability
of the interface to the bottom layer,
is given by
\begin{equation}
\label{BaseReturnProb}
P(\ell) \;=\; \frac{\langle 0 | \, T \,Q^{\ell-2}\, T \,|0\rangle}
{\langle 0| \,T^{\ell}\,|0\rangle} \;\simeq\;
\frac{\langle 0 | \, T \,Q^{\ell-2}\, T \,|0\rangle}{\Lambda^\ell}\,,
\end{equation}
Note that the leftmost column and the topmost row
of the restricted transfer matrix $Q$ are zero,
while all other matrix elements are the same as in Eq.~(\ref{Tmatrix}).
Because of this simple structure, the spectrum of $Q$ is
just the spectrum of $T$ multiplied by $q$ combined with
a zero mode so that the largest eigenvalue of $Q$, which
dominates the matrix product in Eq.~(\ref{BaseReturnProb}),
is $q\Lambda$. In the limit $\ell \to \infty$ we therefore
obtain an exponential distribution of the form
\begin{equation}
P(\ell) \sim q^{\ell} \,.
\end{equation}
Therefore, the average island size scales as
\begin{equation}
\bar{\ell} \simeq -\frac{1}{\ln q} \simeq \frac{1}{\epsilon}
\end{equation}
and diverges at the transition. Since $\rho(0)=1/\bar{\ell}$, this
result is in agreement with Eq.~(\ref{LinearRho}).

\begin{figure}
\includegraphics[width=80mm]{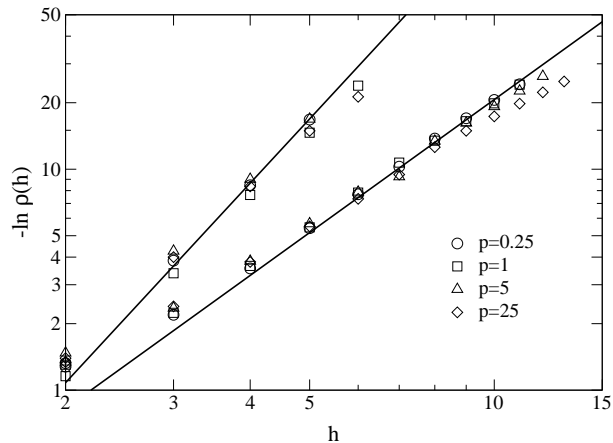}
\caption{\footnotesize
\label{FIGGAUSSIAN}
Numerically determined stationary interface profile of the full model
in 1+1 (lower data points) and 2+1 dimensions (upper data points)
for different values of $p$ slightly below the critical threshold.
The bold lines indicate the slopes $2$ and $3$, respectively.
}
\end{figure}

In order to determine the stationary interface profile $\rho(h)$ in the
limit $h \to \infty$, let us go back to Eq.~(\ref{EigevEquation}).
By assuming that the second and the third term on the left hand side
can be neglected, one finds that
\begin{equation}
\phi_{h} \simeq \frac{1}{\Lambda} \, q^{h-1/2} \,\phi_{h-1}\,,
\end{equation}
with the corresponding asymptotic solution
\begin{equation}
\phi_h \;\sim \; \Lambda^{-h} \, q^{h^2/2}\,.
\end{equation}
Therefore, in the case of detailed balance, the profile of a
bound interface decays as a Gaussian for large values of $h$.
Even for $p\neq 1$, where detailed
balance is violated, numerical simulations (see Fig.~\ref{FIGGAUSSIAN})
suggest that in 1+1 dimensions the entire bound phase is characterized by Gaussian
density profiles.

%
%

\subsection{First-order phase transition for small values of $q_0$}
%
Let us now consider the influence of an attractive force between
substrate and wetting layer by taking $q_0 < q$. Obviously, the
interface is bound to the wall for $q<1$, while for $q>1$ it will
tunnel through the potential barrier. This means that the critical
point $q_c=1$ remains unchanged. However, if $q_0$ is decreased
below a certain threshold~$q_0^*$, the attraction is sufficiently
strong such that the transition becomes first order.

To demonstrate the crossover to a discontinuous transition in the
case of detailed balance, we look for a localized pinned interface
solution at the transition point $q=1$. Assuming an exponential
interface profile
\begin{equation}
\phi_h=z^h \quad \text{ for } h \geq 1
\end{equation}
with some $z<1$, the eigenvalue problem then reduces to three
independent equations
\begin{equation}
\begin{split}
q_0^{-1} \phi_0 + q_0^{-1/2} z &= \Lambda \phi_0\\
q_0^{-1/2} \phi_0 + z + z^2 &= \Lambda z\\
z^{-1} + 1 + z &= \Lambda
\end{split}
\end{equation}
which have the (unnormalized non-negative) solution
\begin{equation}
\phi_0=q_0^{1/2},\qquad
z=\frac{\sqrt{1 + 2 q_0 - 3 q_0^2 }}{2(1-q_0)}-\frac12,\qquad
\Lambda = \frac{z+1}{q_0}\,.
\end{equation}
Consequently, the stationary density of exposed sites
at the bottom layer is
\begin{equation}
\label{rho0}
\rho(0)=\frac{\phi_0^2}{\sum_{h=0}^{\infty}\phi_h^2}
=\frac{q_0}{q_0+\frac{z^2}{1-z^2}}
= \frac{1+q_0-6q_0^2+\sqrt{1 + 2 q_0 - 3 q_0^2 }}
             {2+4q_0-6q_0^2}\,,
\end{equation}
while the densities at higher levels are given by
\begin{equation}
\label{rhoh}
\rho(h)=\frac{\phi_h^2}{\sum_{h=0}^{\infty}\phi_h^2}
=\frac{z^{2h}}{q_0+\frac{z^2}{1-z^2}}\,.
\end{equation}
It turns out that the bottom layer density $\rho(0)$
is positive for $q_0<2/3$ and vanishes
at $q_0=2/3$. For $q_0>2/3$, however,
one has $z>1$, so that the exponential ansatz $\phi_h=z^h$ is no longer
physically meaningful. Hence, in the case of detailed balance, the
transition becomes first order at the tricritical point
\begin{equation}
p=q_c=1\,, \qquad q^*_0 = \frac23\,.
\end{equation}
For $q_0<\frac{2}{3}$, the potential well is deep
enough to bind the critical interface to the wall,
leading to an exponentially decaying interface profile.
For $q_0>\frac{2}{3}$, such a localized solution does not exist
and the transition becomes continuous.

\subsection{Scaling properties near the tricritical point}
%
%
\begin{figure}
\includegraphics[width=80mm]{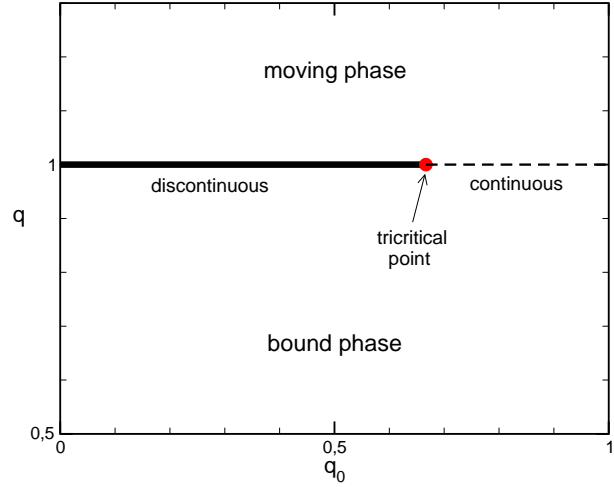}
\caption{\footnotesize
\label{FIGQQ}
Phase diagram for $p=1$ in the $(q,q_0)$-plane.
}
\end{figure}
The phase diagram for $p=1$ is shown in Fig.~\ref{FIGQQ}.
Using the transfer matrix approach, we can show that the
critical behavior along the second order transition line is always
the same as the one discussed in Sec.~\ref{SecondOrderScaling}.
However, in the vicinity of the tricritical point the
scaling properties are different. Moreover, they
depend on the direction from which the tricritical point is
approached.

Let us first consider the case $q=1$, approaching the tricritical
point horizontally from the left along the first-order phase
transition line. Using the expressions~(\ref{rho0})
and~(\ref{rhoh}) one can compute the interface height
\begin{equation}
\bar{h} \;=\; \sum_{h=0}^\infty h \rho(h) \;=\;
\frac{2q_0}{1+q_0-6q_0^2+\sqrt{1+2q_0-3q_0^2}}
\end{equation}
and the squared interface width
\begin{equation}
w^2 \;=\;  \sum_{h=0}^\infty (h-\bar{h})^2 \rho(h) \;=\;
\frac{2\Bigl( 1+4q_0-3q_0^3+(q_0^2-3q_0-1)\sqrt{1+2q_0-3q_0^2}\Bigr)}
     {q_0^2\Bigl(1+3q_0-3\sqrt{1+2q_0-3q_0^2}\Bigr)^2} \,.
\end{equation}
Approaching the tricritical point from the left by increasing $q_0$,
these quantities scale to lowest order in $\delta=q_0^*-q_0$ as
\begin{equation}
\label{ScalingDelta}
\begin{split}
&\bar{h} = w = \frac{1}{6\delta}\\[3mm]
&\rho(0) =  4\delta
\end{split}
\end{equation}
On the other hand,
if the tricritical point is approached vertically keeping
$q_0=q_0^*=2/3$ fixed, a numerical diagonalization of the
transfer matrix suggests that the asymptotic interface profile
crosses over from an exponential to a Gaussian decay.
In this case the height, the width, and the
bottom layer density are found to scale as
\begin{equation}
\label{ScalingEpsilon}
\begin{split}
&\bar{h} \sim w \sim \epsilon^{-1/3}\\
&\rho(0) \sim  \epsilon^{1/3} \,,
\end{split}
\end{equation}
where $\epsilon=1-q$. Moreover, by keeping the boundary
sites fixed at zero height, we can study finite-size scaling
at the tricritical point. Evaluating products of the transfer
matrix numerically we find that
\begin{equation}
\label{SpatialScaling}
\begin{split}
&\bar{h} \sim N^{1/2}\\
&\rho(0) \sim  N^{-1/2} \,.
\end{split}
\end{equation}
\begin{figure}
\includegraphics[width=80mm]{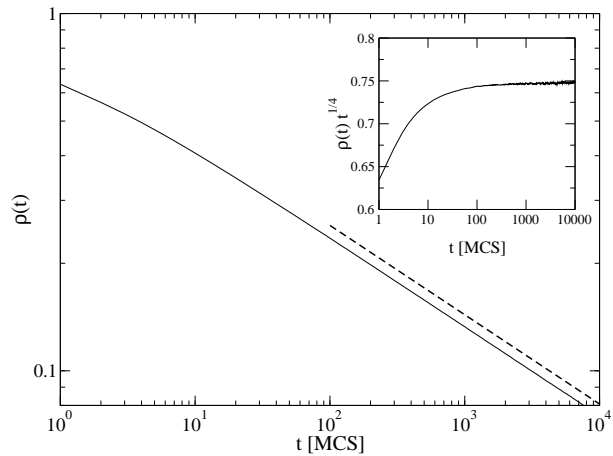}
\caption{\footnotesize
\label{FIGTC}
Decay of the density of sites at zero height at the tricritical point $q=p=1$, $q_0=2/3$.
The slope of the curve tends to $-0.24(1)$, leading to the conjecture that $\rho(0)\sim t^{-1/4}$.
}
\end{figure}
Finally, numerical Monte Carlo simulations at
the tricritical point (see Fig.~\ref{FIGTC})
suggest that an initially flat
interface roughens in such a way that
\begin{equation}
\label{TemporalScaling}
\begin{split}
&\bar{h} \sim t^{1/4} \\
&\rho(0) \sim  t^{-1/4} \,,
\end{split}
\end{equation}
Assuming standard power law scaling all these results can be combined in the
scaling forms
\begin{equation}
\begin{split}
\tilde{h}(\epsilon,\delta,N,t) &= N^{1/2}  F\Bigl(t/N^2,\epsilon N^{3/2},\delta N^{1/2}\Bigr)\\
\rho_0(\epsilon,\delta,N,t)    &= N^{-1/2} G\Bigl(t/N^2,\epsilon N^{3/2},\delta N^{1/2}\Bigr)
\end{split}
\end{equation}
where $F$ and $G$ are scaling functions with an appropriate
asymptotic behavior. Again these scaling forms are
consistent with the critical exponents $z=2, \alpha=1/2$ of the
Edwards-Wilkinson universality class.

\section{Mean field theory for non-equilibrium wetting}
\label{MFSection}
Mean field theories describe a system in an approximate way by
ignoring spatial correlations. For a given model there are many
possible types of mean field theories, depending on the
microscopic level one is trying to describe. Previous mean field
approaches to non-equilibrium wetting considered the average
height as the dynamical variable and studied the mean field
approximation of its dynamics. For example, in a study by Giada
and Marsili~\cite{GiadaMarsili} the KPZ
equation~(\ref{LangevinEquation}) was mapped by a Hopf-Cole
transformation to a Langevin equation with multiplicative noise,
discretizing space and replacing nearest-neighbor interactions by
global couplings. Using a Morse potential with a potential well at
zero height they were able to reproduce second- and first order
transitions as well as phase coexistence. More recently, Santos
{\it et al.}~\cite{Santos} extended these studies, suggesting that
the mean field phase diagram does not change if fluctuations are
taken into account. Moreover, they identified a narrow domain
close to the borderline between phase coexistence region and wet
phase, where the system exhibits spatio-temporal intermittency.

In this Section  we construct mean field equations describing the
temporal evolution of the height distribution~$\psi_n$ for height
$h=n$. This approach is expected to yield more detailed
information on the structure of the interface than previous mean
field theories.

\subsection{Mean field equations}

In order to construct the mean field equations for the height
distribution, let us first consider the case of a freely evolving
interface without a wall. The rate equations describing the
temporal change of~$\psi_n$ consist of several terms corresponding
to different processes. Let us, for example, consider the
probability that an attempted update leads to the desorption of an
atom from the interior of a plateau at level $n$. This probability
is proportional to $\psi_n$, which is the probability to find a
randomly selected site at height $n$. Moreover, it depends on the
heights of the nearest neighbors, which are restricted to take the
values $\{n-1,n,n+1\}$. For simplicity we assume that each site
has two nearest neighbor sites. Clearly one can generalize it to
the case where each site has $2(d-1)$ nearest neighbors. In this
case one may have several types of edge of corner sites, requiring
a larger number of growth rate parameters. To avoid this
complication, we restrict ourselves to the case of two nearest
neighbors. Thus, neglecting spatial correlations, the probability
to find the two nearest neighbor sites at the heights $l$,$m \in
\{n-1,n,n+1\}$ is assumed to be proportional to $\psi_l\psi_m$
divided by $(\psi_{n-1}+\psi_n+\psi_{n+1})^2$. For example, for 
evaporation from a plateau we have $n=l=m$ and thus the contribution 
of the process (\ref{Process3}) to the dynamical equation of $\psi_n$ is
\begin{equation}
- p\frac{\psi_n^3}{(\psi_{n-1}+\psi_n+\psi_{n+1})^2}\,.
\end{equation}
Similarly the loss of probability due to evaporation at the edges (\ref{Process2}) is given by
\begin{equation}
- \frac{2\psi_n^2\psi_{n-1} + \psi_n \psi_{n-1}^2}{(\psi_{n-1}+\psi_n+\psi_{n+1})^2}
\end{equation}
while the depositon process (\ref{Process1}) leads to a loss term of the form
\begin{equation}
- q\frac{\psi_n^3+2\psi_n^2\psi_{n+1}+\psi_n\psi_{n+1}^2}{(\psi_{n-1}+\psi_n+\psi_{n+1})^2}\,.
\end{equation}
In addition, there are corresponding gain contributions at
the neighboring levels such that the total probability is conserved.
Collecting all terms, the mean field equations can be written as
\begin{equation}
\label{MeanFieldEquations}
\frac{d \psi_n}{dt}=A_n - A_{n-1}\,,
\end{equation}
where
\begin{equation}
A_n\;=\;
\frac{p \psi_{n+1}^3 + 2 \psi_n \psi_{n+1}^2 + \psi_n^2 \psi_{n+1}}
{(\psi_{n}+\psi_{n+1}+\psi_{n+2})^2}
- q\,\frac{ \psi_n^3 + 2 \psi_n^2 \psi_{n+1} + \psi_n \psi_{n+1}^2}
{(\psi_{n-1}+\psi_{n}+\psi_{n+1})^2} \,.
\end{equation}
The hard-core wall at the bottom layer can be taken into account by
formally setting $\psi_{-1}=A_{-1}=0$ and replacing $q$ with $q_0$
in the expression for $A_0$, which then can be written as
\begin{equation}
A_0\;=\;
\frac{p \psi_{1}^3 + 2 \psi_0 \psi_{1}^2 + \psi_0^2 \psi_{1}}
{(\psi_{0}+\psi_{1}+\psi_{2})^2}
- q_0\, \psi_0 \,.
\end{equation}
%
%

\subsection{Mean field phase diagram}
%
%
%
\begin{figure}
\includegraphics[width=80mm]{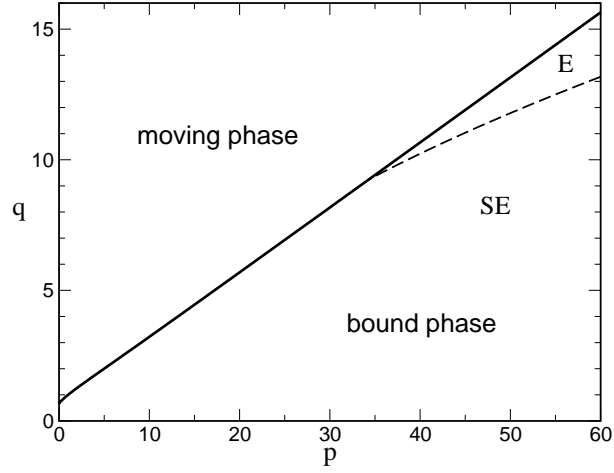}
\caption{\footnotesize
\label{FIGMFPHASEDIAG}
Mean field phase diagram for $q_0=q$. The second-order phase transition line is
represented as a solid line. The bound phase consists of two parts,
where the interface profile $\psi_n$ either decays expontially (E) or
superexponentially (SE) for large $n$.
}
\end{figure}
The main result of the calculations, which will be presented in
detail below, is the mean-field phase diagram shown in
Fig.~\ref{FIGMFPHASEDIAG} for the case $q_0=q$. As in the $1+1$
dimensional model, there is a continuous phase transition from a
bound to a moving phase. In calculating the mean field height
distribution in the moving phase it is found that the interface is
localized around its average height at any given time,
representing a {\em smooth} growing interface. This is expected in
high dimensions, for which mean field represents a reasonable
approximation. Clearly, in $d=1+1$ dimensions this is not the case
as discussed before.

In the bound region, two types of phases were found. In the larger part
of the $p,q$-plane (denoted as {\tt SE} in Fig.~\ref{FIGMFPHASEDIAG}) the
height profile decays superexponentially at large height. In particular
it is found that for large $n$ the profile takes the form
\begin{equation}
\label{Superexp}
\psi_n \sim q^{-n} \, \exp(-2^{n-\alpha})\,,
\end{equation}
where $\alpha$ is a constant. In addition, another region is found
(denoted as {\tt E} in Fig.~\ref{FIGMFPHASEDIAG}), where the height
distribution decays exponentially
\begin{equation}
\label{NormalExp}
\psi_n \sim e^{-an}\,,
\end{equation}
where $a>0$ is a constant.

The superexponential behavior may be understood by considering the
equilibrium case (i.e. $p=1$) in which a $d-1$-dimensional
manifold is attracted by a gravitational force to a hard wall. The
energy of a fluctuation reaching a height $n$ scales as $n^d$,
leading to the height distribution
\begin{equation}
\label{DistribFiniteD} \psi_n \sim e^{-\beta n^d}
\end{equation}
where $\beta >0$ is a constant. This behavior is seems to be valid
in an extended non-equilibrium region of the phase diagram, as
shown in Fig.~\ref{FIGGAUSSIAN}, where numerical simulations in
$d=2+1$ dimensions are presented.

Numerical attempts to find signatures of a possible exponential
phase in $2$ and $3$ dimensions failed since it is very difficult
to obtain a reliable statistics in the tail of the height
distribution, especially in higher dimensions, where finite size
effects become increasingly relevant. The conjecture, that the
exponential phase might be related to the roughening transition of
KPZ interfaces in $d>2$ could not be substantiated by numerical
simulations.

\subsection{Mean field equations}

In the stationary state one has $A_0=0$ and $A_n-A_{n-1}=0$ so
that $A_n$ vanishes for all $n \ge 0$. Therefore, the stationary
mean field equations read
\begin{eqnarray}
0 &=&
p \psi_{1}^3 + 2 \psi_0 \psi_{1}^2 + \psi_0^2 \psi_{1}
- q_0\, \psi_0 (\psi_{0}+\psi_{1}+\psi_{2})^2\,,\\
0 &=&
p \psi_{n+1}^3 + 2 \psi_n\psi_{n+1}^2  + \psi_n^2\psi_{n+1} -
q \psi_n(\psi_n+\psi_{n+1})^2\Bigl(
\frac{\psi_{n}+\psi_{n+1}+\psi_{n+2}}{\psi_{n-1}+\psi_{n}+\psi_{n+1}}
\Bigr)^2\,,
\label{smfe}
\end{eqnarray}
where $n=1,2,\ldots,\infty$.
In order to solve this equation by iteration, it is convenient
to consider quotients of successive densities
\begin{equation}
x_n = \frac{\psi_n}{\psi_{n-1}} \,.
\end{equation}
\begin{figure}
\includegraphics[width=100mm]{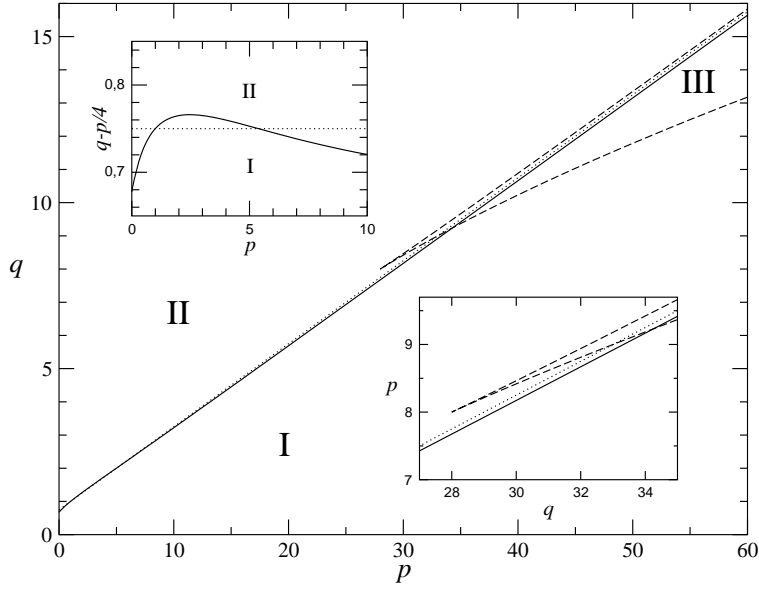}
\caption{\footnotesize
\label{FIGTECH}
Classification of fixed points depending on $p$ and $q$.
In region I (II) there exists only one real fixed point
$x*<1$ ($x*>1$), while  in region III there are three real
fixed points. Regions I and II are separated by the
straight dotted line $q=(p+3)/4$. The phase transition
line (solid line) crosses from region I to region II
and back into region I, as illustrated in the upper inset.
The lower inset zooms the area where the both lines enter
wedge-shaped region III.
}
\end{figure}
\noindent
In terms of these variables, the stationary
mean field equations take the form
\begin{eqnarray}
\label{XIteration0}
0 &=& px_1^3 + 2x_1^2 + x_1 - q_0(1+x_1+x_1x_2)^2 \,, \\
\label{XIteration}
0 &=& p x_{n+1}^3 + 2x_{n+1}^2 + x_{n+1} - q(1+x_{n+1})^2 x_n^2
\Bigl( \frac{1+x_{n+1}+x_{n+1}x_{n+2}}{1+x_n+x_nx_{n+1}} \Bigr)^2 \,.
\end{eqnarray}
The bulk equation (\ref{XIteration}) can be interpreted as an iterative map
$(x_n,x_{n+1})\to (x_{n+1},x_{n+2})$, where
\begin{equation}
\label{Map}
x_{n+2} \equiv f(x_n,x_{n+1}) =  -1-\frac{1}{x_{n+1}} +
\frac{1+x_n+x_nx_{n+1}}{x_n(1+x_{n+1})}
\sqrt{\frac{1+2x_{n+1}+px_{n+1}^2}{q x_{n+1}}}
\end{equation}
for $n=1,2,\ldots\infty$. In addition the initial condition for $x_1,x_2$ is given by
the bottom layer equation
\begin{equation}
\label{BottomCurve}
x_2 = -1-\frac{1}{x_1} + \sqrt{\frac{1+x_1(2+px_1)}{q_0x_1}} \quad .
\end{equation}
%
%

\subsection{Stationary solutions}
In order to evaluate the stationary height distribution for
given $p$ and $q$, one has to look for a point on the line of
possible initial conditions~(\ref{BottomCurve}) and
iterate the map~(\ref{Map}) such that it reaches a
real fixed point with $x^*<1$. This trajectory then corresponds
to a physical height distribution.

To proceed, we first analyze the fixed points of the map.
The fixed point equation
\begin{equation}
\label{FixedPointEquation2}
x\,\Bigl(qx^3 + (2q-p) x^2 + (q-2)x - 1\Bigr) = 0
\end{equation}
has four solutions. Two of the solutions, $x^*_0=0$ and $x^*_1$ are real
in the entire $p,q$-plane, while other two solutions $x^*_2,x^*_3$ may either be both
real or complex conjugate to each other. We denote the region in the $p,q$-plane,
where $x^*_2,x^*_3$ are real, by region III (see Fig.~\ref{FIGTECH}). We further divide the complementary region to III into two regions, I where $x^*_1<1$, and II, where $x^*_1>1$.
Since the bracket in Eq.~(\ref{FixedPointEquation2}) is equal to $-1$ at $x=0$ and positive for
$q>0$ in the limit of large $x$ the fixed point $x_1^*$ has to be positive.

Analyzing the map~(\ref{Map}) with the initial
condition~(\ref{BottomCurve}) we find that the only fixed points, which
correspond to physical height distributions, are either $x^*_0$ or $x^*_2$
when it is real. Here $x_2^*$ is the solution of Eq.~(\ref{FixedPointEquation2})
which satisfies $x_2^* < 1 < x_3^*$ in the region III.
In particular we find that below the transition line, the height distribution
is controlled by the fixed point $x^*_0=0$ in region I and by the fixed
point $x_2^*$ in region III. Details of the analysis, which led to this
result, are given in the Appendix.

We now study the stationary height profile in the two regions.
In region I the map flows to the hyperbolic fixed point $x^*_0=0$
along its stable trajectory. Expanding the map for small values of $x$ to
lowest order, this stable manifold is described by the nonlinear relation
\begin{equation}
\label{NonlinearFixedPoint}
x_{n+1} \simeq q \, x_n^2\,,
\end{equation}
in the limit $x \to 0$. Along this manifold, the
map approaches the fixed point  $x^*_0=0$ super-exponentially as
\begin{equation}
x_n \simeq \frac{1}{q}\exp({-2^{n-\alpha}})
\end{equation}
yielding the height profile in Eq.~(\ref{Superexp}).

In region III stationary solutions are controlled by the
fixed point $x_2^*>0$ (see Appendix). Linearizing the map around this
fixed point one obtains an exponential behavior
\begin{equation}
x_n \sim e^{-an}\,,
\end{equation}
where $a>0$ is a constant, leading to the exponential
height profile~(\ref{NormalExp}).

To complete the analysis of the phase diagram, one has to locate
the wetting transition line. This is done by simulating the
dynamical mean field equations~(\ref{MeanFieldEquations}) and
determining the point from where on the interface detaches. This
analysis leads to the transition line shown in
Fig.~\ref{FIGMFPHASEDIAG}.

So far all mean field results were obtained for $q_0=q$, i.e.,
without an attractive force between the substrate and the wetting
layer. Lowering $q_0$ changes the bottom layer
equation~(\ref{BottomCurve}) and thereby the possible starting
points of the iteration. As shown in the Appendix, the mean field
approximation reproduces the phenomenon of phase coexistence.
However, unlike in the model in $1+1$ dimensions, it emerges
everywhere along the phase transition line as soon as $q_0<q$;
hence, there is no tricritical point. Moreover, in the limit $q_0
\to 0$ the threshold for the growth rate $q$, where the interface
detaches, tends to infinity. Therefore, the region of phase
coexistence is not bounded from above as in the $1+1$ dimensional
model.

\section{Conclusions}
\label{SecConcl}

In this paper we have given a detailed account of a recently introduced
solid-on-solid model for non-equilbrium wetting, which is defined
in the spirit of a KPZ equation in a potential. Introducing a
hard-core wall at zero height, the model exhibits a contininuous
wetting transition from a bound to a moving phase. The model is
controlled by two paramters $p$ and $q$, which effectively determine
the asympotic slope of the potential and the coefficient of the
nonlinear term in the KPZ equation.

For $p=1$ the dynamical rules of the model obey detailed balance.
In this case the stationary distribution of a bound interface is
given by a Boltzmann ensemble, which allows one to derive various
quantities exactly. Moving away from this line, the model crosses
over to a non-equilibrium behavior, that is characterized by
different critical properties. The model can be generalized
further by including an attractive interaction between the
substrate and the wetting layer. If this force is strong enough it
may turn the continuous transition into a discontinuous one.
Moreover, the bound and the moving phase may coexist in regions
where the coeffient of the nonlinear term in the KPZ equation is
negative.

In order to assess the behavior of the model in higher dimensions,
we have proposed a mean field approximation, which is based on rate
equations for the densities at different heights. The phase diagram
turns out to be surprisingly rich. It turns out that the mean field
approximation reproduces the properties of the original model.
However, in the moving phase the interface remains smooth as it
is expected for KPZ-type growth in higher dimensions.

As a new feature, the mean field approximation predicts the existence
of two different regions in the bound phase. In one of these regions
the interace profile decays superexponentially with increasing height
while in the other region an exponential decay is observed. This
can be explained by classifying the fixed points of an iterative
map for quotients of the densities. The question to what extent
this crossover from superexponential to exponential profiles can
be observed in the full model is still open.


\noindent
{\bf Acknowledgments}\\
The support of the Israel Science Foundation (ISF) and the Einstein
center of the Minerva foundation is gratefully acknowledged. The
simulations were partly performed on the ALiCE parallel computer
at the IAI in Wuppertal. H.H would like to thank the INFN and the
Weizmann Institute for hospitality, where parts of this work have
been done.

\newpage

\appendix
\section{Stationary solutions of the mean-field equations}
\label{appendix}
%
%
\subsection{Fixed points}
%
%
\noindent Besides $x_0^*=0$, Eq.~(\ref{FixedPointEquation2}) has
three other fixed points which are the roots of the polynomial
equation
\begin{equation}
\label{FixedPointEquation}
qx^3 + (2q-p) x^2 + (q-2)x - 1 = 0 \,.
\end{equation}
As shown in Fig.~\ref{FIGTECH}, the $p,q$-plane can be divided into three
different regions. In regions I and II one fixed point is real and two of them are
complex conjugate, while in region III all fixed points are real. The
boundary of region III is characterized by the existence of a two-fold
degenerate fixed point. Comparing Eq.~(\ref{FixedPointEquation})
with a polynomial of the form $(x-a)^2(x-b)=0$, the boundary of region III
can be given in a parameter representation by
\begin{equation}
\label{ParameterRepresentation}
q = \frac{2}{a-a^2} \,, \qquad p = \frac{1+3a+4a^2}{a^2-a^3}
\,, \qquad b=\frac{1-a}{2a} \,,
\end{equation}
where $0<a<1$. The boundary has the form of a wedge,
shown as a dashed line in Fig.~\ref{FIGTECH}. The
lower branch for $0<a<1/2$ and the upper branch for $1/2<a<1$ terminate
in the triple point $(p,q)=(8,28)$, where
Eq.~(\ref{FixedPointEquation}) has a three-fold degenerate fixed point
$x^*=1/2$.

Physically meaningful stationary solutions of the iterative map
must start from a point on the line given by
Eq.~(\ref{XIteration0}) and have to flow towards a real fixed
point with $0 \leq x^* < 1$. For this reason we divided the
complement of region III into two parts, namely, region I with
$x_1^*>1$ and region II with $x_1^*<1$. Both regions are separated
by a straight line $q=(p+3)/4$, where $x_1^*=1$.

%
\subsection{Stationary solutions in regions I and II}
%
We first show that in regions I and II a physically meaningful
stationary solution always flows to the fixed point $x_0^*=0$.
To this end we show that the other real fixed point $x_1^*$ is either
larger than $1$ or unstable.

\begin{figure}
\includegraphics[width=160mm]{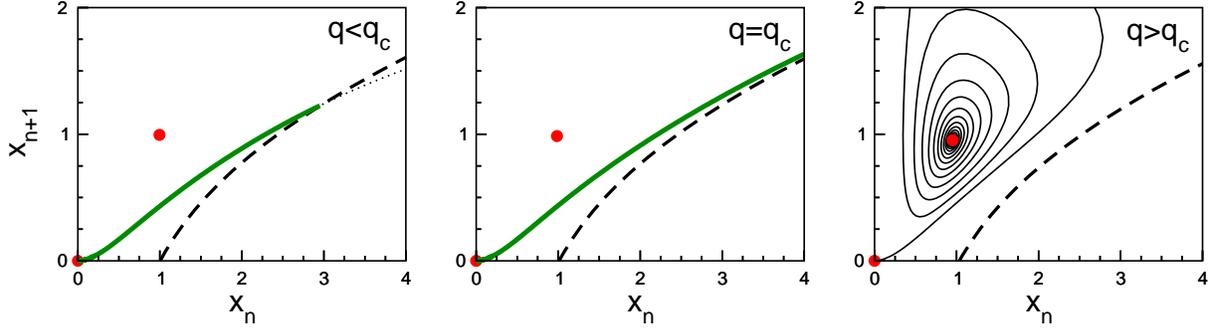}
\caption{\footnotesize
\label{FIGOUTSIDE}
Stationary solutions in region I:
Superstable manifold of the fixed point $x^*_0=0$ for $p=2$
and different values of $q$ in regions I and II.
For $q<q_c$ the manifold (bold line) interesects with
the with the dashed line of possible initial conditions (\ref{BottomCurve}),
representing a stationary solution in the bound phase, where
the bottom layer density $\psi_0$ is positive. Approaching
$q_c$ this intersection point moves to infinity and $\psi_0$
tends to zero. For $q>q_c$ the
superstable manifold originates in the other real fixed
point so that no stationary solution exists. Similar graphs
are also obtained for other values of $p$ in regions I and II.
}
\end{figure}

As shown in the upper inset of Fig.~\ref{FIGTECH}, the phase
transition line starts at $p=0$ in region I, crosses into region
II at the point $q=p=1$ and then crosses back into region I at the
point $(p,q)\approx(5.380,2.095)$. Obviously the fixed point
$x_1^*$ can only be physically meaningful between these two
crossing points, where $x_1^*<1$. However, in this interval
$x_1^*$ turns out to be unstable. To demonstrate this point we
consider the eigenvalues of the Jacobian of the map
$x_{n+2}=f(x_n,x_{n+1})$ in Eq.~(\ref{Map})
\begin{equation}
\label{Eigenvalues}
\lambda_{1,2} = \frac12 \left.\left(
\frac{\partial f}{\partial x_{n+1}} \pm
\sqrt{\Bigl(\frac{\partial f}{\partial x_{n+1}}\Bigr)^2
+4\frac{\partial f}{\partial x_n}}\,
\right)\right|_{x_n=x_{n+1}=x^*}\,.
\end{equation}
Using Eq.~(\ref{FixedPointEquation}) the partial derivatives can be expressed as
\begin{equation}
\begin{split}
\left.\frac{\partial f}{\partial x_n}\right|_{x_n=x_{n+1}=x^*} &=
-\frac{1}{{x^*}^2} \,,\\
\left.\frac{\partial f}{\partial x_{n+1}}\right|_{x_n=x_{n+1}=x^*}  &= \frac
{q{x^*}^4+2q{x^*}^3+2(q-1){x^*}^2+(3q-2){x^*}-2}
{2q\,{x^*}^3({x^*}+1)} \,.
\end{split}
\end{equation}
Along the dotted line in Fig.~\ref{FIGTECH}, where $x^*=1$, the
two eigenvalues
\begin{equation}
\left.\lambda_{1,2}\right|_{x^*=1} \;=\; 1-\frac{1}{4q}(3 \pm \sqrt{9-24q})
\end{equation}
are complex conjugate in the interval between the two
crossing points $1<q<5.380$. As can be verified numerically, they
are also complex conjugate in a neighborhood of this line,
which includes the phase transition line.
Since the determinant of the Jacobian
\begin{equation}
\lambda_1\lambda_2 = \frac{1}{{x^*}^2}
\end{equation}
is larger than $1$ for $x^*<1$, the fixed point $x_1^*$ is found to be unstable.
Thus we can conclude that physically meaningful stationary solutions
in regions I and II are always controlled by the fixed point $x_0^*=0$.

Since for $x^*\to 0$ the two eigenvalues tend to $\lambda_1=0$ and
$\lambda_2=-\infty$, the fixed point $x_0^*=0$ is nonlinear and
hyperbolic. It has superstable manifold which in the limit of
$x \to 0$ is given by $x_{n+1}=qx_n^2$. This manifold is shown in
Fig.~\ref{FIGOUTSIDE} for different values of $q$ below, at, and
above the critical point. As it can be seen, a stationary solution
exists if the manifold intersects the line of possible
initial conditions~(\ref{BottomCurve}).

%
\subsection{Stationary solutions in region III}
%

%
%
\begin{figure}
\includegraphics[width=160mm]{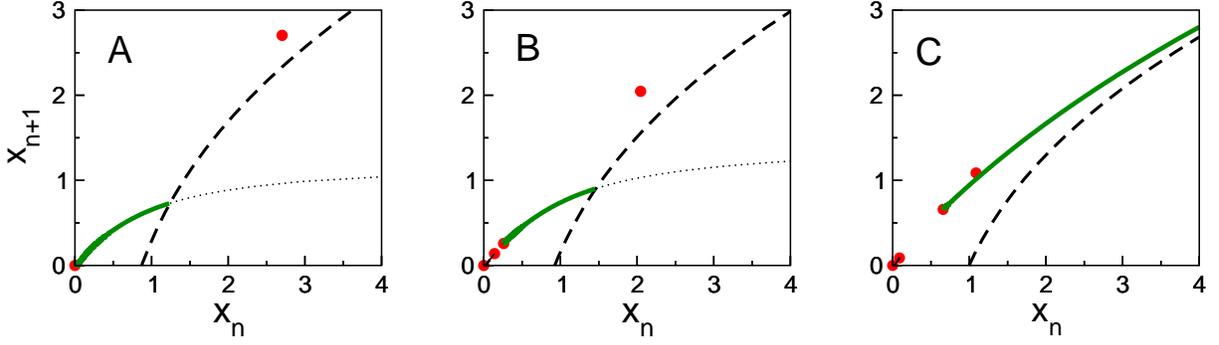}
\caption{\footnotesize
\label{FIGINSIDE}
Fixed points (bullets) and possible
stationary solutions (bold lines) for $p=60$ inside for
(A) $q=12.0$, (B) $q=13.5$, and (C) $q=q_c\simeq 15.644$.
The figure is explained in the text.
}
\end{figure}

In the wedge-shaped region III, there are three real fixed points 
$x_1^*<x_2^*<x_3^*$.
The first one is smaller than $1$ and unstable, while the
second one is smaller than $1$ and hyperbolic. The properties of $x_3^*$
depend on $p$ and $q$. Below the dotted line in Fig.~\ref{FIGTECH} it is
larger than $1$ and stable, while it is smaller than $1$ and unstable above.

Fig.~\ref{FIGINSIDE} illustrates typical situations at
$p=60$ for various values of $q$, crossing the region III from
bottom to top. Below the wedge in region I (panel A),
the stationary solution can be constructed in
the same way as before. Entering region III from below (panel B),
two new real fixed points $x_1^*<x_2^*<1$ emerge,
which are fully unstable and hyperbolic, respectively.
The superstable manifold of the fixed point $x_0^*=0$
now originates in $x_1^*$, while it is the linearly stable manifold
of the hyperbolic fixed point $x_2^*$ which intersects the
dashed line of possible initial conditions~(\ref{BottomCurve}).
As before, the intersection point moves
continuously to infinity as $q$ approaches the critical point (panel C).
Thus the unbinding transition manifests itself in the same
way as in the regions I and II, the only difference being that
the physically relevant stable manifold is now controlled by the hyperbolic fixed
point $x_2>0$ instead of $x_0^*=0$. The linear stability of $x_2^*$ is
responsible for the purely exponential profile observed in region $E$.

%
%
\subsection{Phase coexistence in regions I and II}
\begin{figure}
\includegraphics[width=120mm]{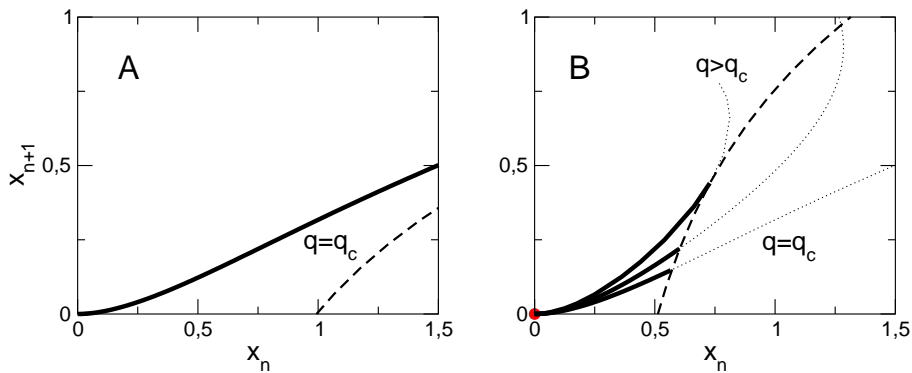}
\caption{\footnotesize
\label{FIGMFCOEX}
Phase coexistence in the mean field approximation
for p=0.8 (see text).
}
\end{figure}

%
%
Lowering $q_0$ changes the line of possible initial
conditions~(\ref{BottomCurve}), while the stable
manifold of the fixed point $x_0^*=0$ remains the same.
The typical situation for $p=0.8$ is shown in Fig.~\ref{FIGMFCOEX}.
The left panel shows the stable manifold (solid line)
and the bottom layer equation (dashed line) without attractive force
at the critical point $q_0=q=q_c\simeq 0.943$.
Both curves approach each other smoothly and intersect at infinity
so that the transition is continuous. The right panel
shows the same situation in the presence
of an attractive force for $q_0=0.5$. Accordingly, the effect
of lowering $q_0$ is to move the dashed line of possible intial
conditions upward such that it intersects the
critical stable manifold at a certain finite point. This means
that the bottom layer density $\psi_0$ is finite
at the critical point, making the transition first order.
Moreover, a stationary solution still exists
even if $q$ is increased, proving the possibility
of phase coexistence in the mean field equations.
Increasing $q$ beyond a certain threshold, a stationary solution no 
longer exists.
This defines the upper boundary of the phase coexistence region
in the mean field phase diagram.

%
%


\end{document}